\documentclass[english,preprint,nofootinbib,byrevtex]{revtex4}
\usepackage[T1]{fontenc}
\usepackage[latin1]{inputenc}
\usepackage{amsmath}
\usepackage{graphicx}
\usepackage{amssymb}

\makeatletter

\providecommand{\tabularnewline}{\\}

\usepackage{graphicx}
\providecommand{\tabularnewline}{\\}

\usepackage{slashed}
\usepackage{bm}
\usepackage{mathrsfs}

\newcommand{\ssst}[1]{\scriptscriptstyle{#1}}

\newcommand{\bra}[1]{\langle #1 \vert}
\newcommand{\ket}[1]{\vert #1 \rangle}

\newcommand{\mN}{m_{\ssst{N}}}

\newcommand{\gpiNN}{g_{\pi\ssst{NN}}}
\newcommand{\getaNN}{g_{\eta\ssst{NN}}}
\newcommand{\grhoNN}{g_{\rho\ssst{NN}}}
\newcommand{\gomegaNN}{g_{\omega\ssst{NN}}}

\newcommand{\muS}{\mu_{\ssst{S}}}
\newcommand{\muV}{\mu_{\ssst{V}}}
\newcommand{\chiS}{\chi_{\ssst{S}}}
\newcommand{\chiV}{\chi_{\ssst{V}}}
\newcommand{\kappaS}{\kappa_{\ssst{S}}}
\newcommand{\kappaV}{\kappa_{\ssst{V}}}

\newcommand{\PVTV}{\slashed{P}\slashed{T}}
\newcommand{\HPVTV}{H_{\ssst{\slashed{P}\slashed{T}}}}

\allowdisplaybreaks

\usepackage{babel}
\makeatother
\begin{document}

\preprint{KVI-1660}

\title{$P$- and $T$-odd two-nucleon interaction and the deuteron electric
dipole moment}

\author{C.-P. Liu}

\email{liu@KVI.nl}

\affiliation{Theory Group, Kernfysisch Versneller Instituut, University of Groningen,
\\
 Zernikelaan 25, NL-9747 AA Groningen, The Netherlands}

\author{R.G.E. Timmermans}

\email{timmermans@KVI.nl}

\affiliation{Theory Group, Kernfysisch Versneller Instituut, University of Groningen,
\\
 Zernikelaan 25, NL-9747 AA Groningen, The Netherlands}

\date{\today{}}

\begin{abstract}
The nuclear physics relevant to the electric dipole moment (EDM) of
the deuteron is addressed. The general operator structure of the $P$-
and $T$-odd nucleon-nucleon interaction is discussed and applied
to the two-body contributions of the deuteron EDM, which can be calculated
in terms of $P$- and $T$-odd meson-nucleon coupling constants with
only small model dependence. The one-body contributions, the EDMs
of the proton and the neutron, are evaluated within the same framework.
Although the total theoretical uncertainties are sizable, we conclude
that, compared to the neutron, the deuteron EDM is competitive in
terms of sensitivity to $C\! P$ violation, and complementary with
respect to the microscopic sources of $C\! P$ violation that can
be probed. 
\end{abstract}
\maketitle

\section{Introduction}

In the field of particle physics an atomic physics quantity plays
a privileged role: the electric dipole moment (EDM), which violates
parity ($P$) conservation and time reversal ($T$, or equivalently
$C\! P$) invariance. The Standard Model predicts values for EDMs
that are much too small to be detected in the foreseeable future,
and hence a nonzero EDM is an unambiguous signal of a new source of
$C\! P$ violation~\cite{Barr:1992dq,Khriplovich:1997}.

Over the years, many experiments have searched with increasing precision
for a nonzero EDM. The most sensitive experiments measure the precession
frequency of the spin for neutral systems, such as the neutron or
an atom, in a strong electric field. The limit on the EDM of the neutron,
in particular, has been improved spectacularly over the years~\cite{Ramsey:1991gw}.
The most precise value obtained so far is $d_{n}=(-1.0\pm3.6)\times10^{-26}$
$e$--cm~\cite{Harris:1999jx}. New experiments using high-density
ultracold neutron sources are being set up to target the precision
level of $10^{-27}$ to $10^{-28}$ $e$--cm at LANL (LANSCE), PSI,
ILL, and Munich (FRM-II).

Limits on the EDMs of charged particles~\cite{Sandars:2001nq}, such
as the electron and the proton, has so far been derived from experiments
with selected neutral atoms (and molecules). The best limit for an
EDM has been obtained for the $^{199}$Hg mercury atom~\cite{Romalis:2000mg},
for which $d_{\textrm{Hg}}=(-1.06\pm0.49\pm0.40)\times10^{-28}$ $e$--cm
was measured. In such a closed-shell atom with paired electron spins,
the EDM of the atom arises mainly from the EDM of unpaired nucleons
and from $T$-odd interactions within the nucleus. For this type of
experiments with neutral atoms, the EDM signal is severely suppressed
due to the screening of the applied external electric field by the
atomic electrons, a general result known in the literature as Schiff's
theorem~\cite{Purcell:1950,Schiff:1963,Engel:1999np}.

Recently, a new highly sensitive method has been proposed to directly
measure the EDMs of charged particles, such as the muon or ions, in
a magnetic storage ring~\cite{Semertzidis:2003iq,Farley:2003wt}.
The method evades the suppression of the EDM signal due to Schiff's
theorem, and works for systems with a small magnetic anomaly. An experiment
using this method has been proposed to measure the EDM of the deuteron
at the $10^{-27}$ $e$--cm level~\cite{-d_EDM}. From a theoretical
point of view, the deuteron is especially attractive, because it is
the simplest system in which the $P$-odd, $T$-odd ($\PVTV$) nucleon-nucleon
($N\! N$) interaction contributes to the EDM. Moreover, the deuteron
properties are well understood~\cite{deSwart:1995ui}, so reliable
and precise calculations are possible.

It is our goal in this paper to address the nuclear physics part of
the deuteron EDM calculation, and to compare the result to the EDM
of the neutron (and proton) evaluated within the same framework. The
framework is designed so that our results for the nucleon EDM and
the $\PVTV$ $N\! N$ interaction would be suitable, when combined
with a realistic strong $N\! N$ interaction, as a starting point
for a microscopic calculation of the EDM of more complex systems,
such as the mercury atom.

This paper is organized as follows. In Section II we construct the
general operator structure of the $P$- and $T$-odd $N\! N$ interaction
and from it derive the potential in terms of strong and $\PVTV$ meson-nucleon
coupling constants. In Section III, we used this $\PVTV$ potential
in combination with modern $N\! N$ potential models to evaluate the
two-body (polarization and exchange) contributions to the deuteron
EDM. The one-body contributions, \textit{i.e.} the EDMs of the proton
and the neutron, are calculated within the same framework. Finally,
the results are discussed and conclusions are drawn in Section IV.
In the Appendix we discuss and evaluate the, also $P$- and $T$-odd,
magnetic quadrupole moment (MQM) of the deuteron.

\section{P- and T-odd Two-Nucleon Interaction}

By contracting two Dirac bilinear covariants containing at most one
derivative, the $P$-odd, $T$-odd, and $C$-even (hence still $C\! PT$-even)
contact $N\! N$ interaction can be constructed from ($i$) the scalar-pseudoscalar
(S--PS) combination, $\bar{N}\! N\times\bar{N}i\gamma_{5}N$, and
($ii$) the vector--pseudovector (V--PV) combination, $\bar{N}\gamma^{\mu}N\times\bar{N}\tensor{\partial}_{\mu}\gamma_{5}N$
\cite{Fischler:1992ha}. The tensor--pseudotensor (T--PT) combination,
$\bar{N}\sigma^{\mu\nu}N\times\bar{N}\sigma_{\mu\nu}\gamma_{5}N$,
also qualifies these symmetry considerations, however, it is equivalent
to the S--PS one by a Fierz transformation.

Using the nonrelativistic (NR) reduction and writing out the isospin
structure explicitly, the most general form of the low-energy, $P$-
and $T$-odd ($\PVTV$), zero-range (ZR) $N\! N$ interaction, $\HPVTV^{\ssst{\mathrm{{(ZR)}}}}$,
can be expressed, in configuration space, as \begin{eqnarray}
\HPVTV^{\ssst{\mathrm{{(ZR)}}}} & = & \frac{1}{2\mN}\bigg\{\big(c_{1}+d_{1})\,\bm\sigma_{-}+(c_{2}+d_{2})\,\bm\tau_{1}\cdot\bm\tau_{2}\,\bm\sigma_{-}+(c_{3}+d_{3})\,\tau_{+}^{z}\,\bm\sigma_{-}+(c_{4}+d_{4})\,\tau_{-}^{z}\,\bm\sigma_{+}\nonumber \\
 &  & +(c_{5}+d_{5})\,(3\,\tau_{1}^{z}\,\tau_{2}^{z}-\bm\tau_{1}\cdot\bm\tau_{2})\,\bm\sigma_{-}\bigg\}\cdot\bm\nabla\,\delta^{(3)}(\bm r)\,,\label{eq:HPVTV-contact}\end{eqnarray}
 where $\bm\sigma_{\pm}\equiv\bm\sigma_{1}\pm\bm\sigma_{2}$ and $\bm\tau_{\pm}\equiv\bm\tau_{1}\pm\bm\tau_{2}$.%
\footnote{We note that this most general NR form containing five independent
isospin-spin operators has already been pointed out in Ref. \cite{Herczeg:1966}.%
} Terms involving the isospin operator $i\,(\bm\tau_{1}\times\bm\tau_{2})^{z}$,
even though they conserve charge, are ruled out since they are $C$-odd.
The dimensionful coupling constants $c_{i}$ and $d_{i}$ ($i=1,...,5$)
each correspond to a unique isospin-spin-spatial operator in the S--PS
and V--PV parts, respectively. These constants are the quantities
that experiments such as nuclear EDM measurements can hopefully constrain,
and thus predictions from different models of $C\! P$ violation could
be tested against with.

At first sight, it seems that the introduction of the $d_{i}$'s is
redundant because the V--PV form has exactly the same NR limit as
its S--PS counterpart, a point which has been made in Ref.~\cite{Fischler:1992ha}.
Therefore, as long as one works strictly in the context of contact
interactions, \textit{e.g.} {}``pionless'' effective field theory,
only five coupling constants are needed to fit to experiments. However,
there are several reasons to justify this larger set, especially when
one goes beyond the ZR limit with several energy scales involved.

First, when one tries to connect the experimental constraints to underlying
theoretical models, it is still necessary to make the distinction
between the S--PS and V--PV sectors. Because of different nucleon
dynamics involved, the separation and comparison of these two sectors
are of interest.

Second, if one wants to keep the pions, as the lightest mesons, explicitly
and model the long-range (LR) interaction through one-pion exchange
(see \textit{e.g.} Refs.~\cite{Barton:1961eg,Haxton:1983dq,Herczeg:1987gp}),
a scale separation defined by the pion mass naturally occurs. In this
case, one has in total eight independent coupling constants: five
in the ZR potential which is a result of integrating out all degrees
of freedom except the pions, and three $\PVTV$ pion-nucleon coupling
constants (see below, Eq. (\ref{eq:L-PVTV})) which describe the LR
potential.%
\footnote{The three $\PVTV$ $\pi N\! N$ couplings were first pointed out by
Barton~\cite{Barton:1961eg}. However, since the concern then was
parity violation, these couplings were only picked up later when interest
in nuclear $C\! P$ violation built up.%
} This possible scale difference between the S--PS and V--PV sectors
is not manifest in the ZR limit.

The third and more practical reason is that we are going to adopt
a {}``hybrid'' approach for the $N\! N$ dynamics which takes advantage
of existing high-quality strong $N\! N$ potentials and use perturbation
theory based on operators constructed in the spirit of effective field
theory (EFT). In such a framework, it is necessary to smear out the
contact interactions. The physical guideline is to take the delta
function as a limit of the mass$^{2}$-weighted Yukawa function, $m_{x}^{2}\,\mathcal{Y}_{x}(r)=m_{x}^{2}\, e^{-m_{x}r}/(4\pi r)$,
when the exchanged boson is taken to be extremely massive: \begin{equation}
\lim_{m_{x}\rightarrow\infty}m_{x}^{2}\,\frac{e^{-m_{x}\, r}}{4\pi r}=\lim_{m_{x}\rightarrow\infty}\textrm{{F.T.}}\left[\frac{m_{x}^{2}}{\bm q^{2}+m_{x}^{2}}\right]=\delta^{(3)}(\bm r)\,,\end{equation}
 where {}``F.T.'' stands for Fourier transform. As suggested above,
allowing different mass scales for the S--PS and V--PV sectors then
leads to the most general $\HPVTV$ in terms of ten independent operators.

Although the choices of the mass parameters for the Yukawa functions
are arbitrary in the sense of fitting the coupling constants, the
mass spectrum of low-lying mesons provides an intuitive choice and
suggests a connection between $\HPVTV$ thus constructed and the one-meson
exchange scheme. Besides the one-pion exchange ($J^{P}=0^{-}$, $m_{\pi}=140$
MeV) often adopted in the literature, the contribution from $\eta$
($J^{P}=0^{-}$, $m_{\eta}=550$ MeV)~\cite{Gudkov:1993yc}, and
from $\rho$ and $\omega$ ($J^{P}=1^{-}$, $m_{\rho,\omega}=770,780$
MeV)~\cite{Towner:1994qe} have also been considered in various works.
We will show that a one-meson exchange scheme containing $\pi$, $\eta$,
$\rho$, and $\omega$ produces the same general operator structure
as the ZR scheme. (The isoscalar-scalar meson $\varepsilon$ or {}``$\sigma$'',
with a $\PVTV$ coupling of type $\bar{N}i\gamma_{5}\sigma N$, leads
to the same operator structure as the $\eta$ meson, and its contribution
would be effectively subsumed in the coupling $\bar{G}_{\eta}^{(0)}$.)

The strong and $\PVTV$ meson-nucleon interaction Lagrangian densities,
$\mathcal{L}_{\ssst{S}}$ and $\mathcal{L}_{\ssst{\PVTV}}$, are%
\footnote{The choice of pseudoscalar coupling for the pion field in Eq.~(\ref{eq:L-strong})
is traditional in the EDM literature. In order to have manifest chiral
symmetry, pseudovector (derivative) coupling is of course preferred.
However, the results for the two-body contributions and for the leading
one-body contribution (the chiral logarithm) would be equivalent.%
} \begin{eqnarray}
\mathcal{L}_{\ssst{S}} & = & \gpiNN\bar{N}i\gamma_{5}\bm\tau\cdot\bm\pi N\nonumber \\
 &  & +\getaNN\bar{N}i\gamma_{5}\eta N\nonumber \\
 &  & -\grhoNN\bar{N}\Big(\gamma^{\mu}-i\frac{\chiV}{2\mN}\sigma^{\mu\nu}q_{\nu}\Big)\,\bm\tau\cdot\bm\rho_{\mu}N\nonumber \\
 &  & -\gomegaNN\bar{N}\Big(\gamma_{\mu}-i\frac{\chiS}{2\mN}\sigma^{\mu\nu}q_{\nu}\Big)\omega_{\mu}N\,,\label{eq:L-strong}\end{eqnarray}
 and \begin{eqnarray}
\mathcal{L}_{\ssst{\PVTV}} & = & \bar{N}\big(\bar{g}_{\pi}^{(0)}\bm\tau\cdot\bm\pi+\bar{g}_{\pi}^{(1)}\pi^{0}+\bar{g}_{\pi}^{(2)}(3\tau^{z}\pi^{0}-\bm\tau\cdot\bm\pi)\big)N\nonumber \\
 &  & +\bar{N}\big(\bar{g}_{\eta}^{(0)}\eta+\bar{g}_{\eta}^{(1)}\tau^{z}\eta\big)N\nonumber \\
 &  & +\bar{N}\frac{1}{2\mN}\big(\bar{g}_{\rho}^{(0)}\bm\tau\cdot\bm\rho_{\mu}+\bar{g}_{\rho}^{(1)}\rho_{\mu}^{0}+\bar{g}_{\rho}^{(2)}(3\tau^{z}\rho_{\mu}^{0}-\bm\tau\cdot\bm\rho_{\mu})\big)\sigma^{\mu\nu}q_{\nu}\gamma_{5}N\nonumber \\
 &  & +\bar{N}\frac{1}{2\mN}\big(\bar{g}_{\omega}^{(0)}\omega_{\mu}+\bar{g}_{\omega}^{(1)}\tau^{z}\omega_{\mu})\sigma^{\mu\nu}q_{\nu}\gamma_{5}N\,.\label{eq:L-PVTV}\end{eqnarray}
 The $g_{\ssst{XNN}}$'s are the strong $X\! N\! N$ coupling constants
for which we will adopt the values: $\gpiNN=13.07$~\cite{Stoks:1993ja,Rentmeester:1999vw},
$\getaNN=2.24$~\cite{Tiator:1994et}, $\grhoNN=2.75$~\cite{Hohler:1975ht}
and $\gomegaNN=8.25$.%
\footnote{We use the prediction~\cite{deSwart:1993te} $\gomegaNN^{2}=9\grhoNN^{2}$
to infer $\gomegaNN$ from $\grhoNN^{2}/4\pi=0.6$ given in Ref.~\cite{Hohler:1975ht}.%
} The $g_{\ssst{X}}^{(i)}$'s are the $\PVTV$ ones with the superscript
$i=0,1,2$ denoting the corresponding isospin content.%
\footnote{We use the Bjorken-Drell metric and special attention should be paid
to the definition of $\gamma_{5}=\left(\begin{array}{cc}
0 & I\\
I & 0\end{array}\right)$ and any coupling constant associated with it. Relative to the Pauli
metric a sign difference due to $\gamma_{5}$ should be kept in mind.%
} $\chiV$ and $\chiS$ are the ratios of the tensor to vector coupling
constant for $\rho$ and $\omega$ respectively; when vector-meson
dominance (VMD)~\cite{Sakurai:1969} is assumed, they are equal to
the electromagnetic counterparts, \textit{i.e.} $\kappaV=3.70$ and
$\kappaS=-0.12$. The tensor structure $\bar{N}\sigma^{\mu\nu}q_{\nu}\gamma_{5}N$
in Eq.~(\ref{eq:L-PVTV}), where $q_{\nu}=p_{\nu}-p'_{\nu}$, is
equivalent to the PV structure $\bar{N}\tensor{\partial}_{\mu}\gamma_{5}N$
by a Gordon decomposition.

Evaluating all one-meson exchange diagrams with one strong and one
$\PVTV$ vertex, the NR potential, $\HPVTV$, is found to be \begin{eqnarray}
\HPVTV & = & \frac{1}{2\mN}\bigg\{\bm\sigma_{-}\cdot\bm\nabla(\bar{G}_{\eta}^{(0)}\,\mathcal{Y}_{\eta}(r)-\bar{G}_{\omega}^{(0)}\,\mathcal{Y}_{\omega}(r))\nonumber \\
 &  & +\bm\tau_{1}\cdot\bm\tau_{2}\,\bm\sigma_{-}\cdot\bm\nabla(\bar{G}_{\pi}^{(0)}\,\mathcal{Y}_{\pi}(r)-\bar{G}_{\rho}^{(0)}\,\mathcal{Y}_{\rho}(r))\nonumber \\
 &  & +\tau_{+}^{z}\,\bm\sigma_{-}\cdot\bm\nabla[\frac{1}{2}(\bar{G}_{\pi}^{(1)}\,\mathcal{Y}_{\pi}(r)-\bar{G}_{\eta}^{(1)}\,\mathcal{Y}_{\eta}(r))-\frac{1}{2}(\bar{G}_{\rho}^{(1)}\,\mathcal{Y}_{\rho}(r)+\bar{G}_{\omega}^{(1)}\,\mathcal{Y}_{\omega}(r))]\nonumber \\
 &  & +\tau_{-}^{z}\,\bm\sigma_{+}\cdot\bm\nabla[\frac{1}{2}(\bar{G}_{\pi}^{(1)}\,\mathcal{Y}_{\pi}(r)+\bar{G}_{\eta}^{(1)}\,\mathcal{Y}_{\eta}(r))-\frac{1}{2}(-\bar{G}_{\rho}^{(1)}\,\mathcal{Y}_{\rho}(r)+\bar{G}_{\omega}^{(1)}\,\mathcal{Y}_{\omega}(r))]\nonumber \\
 &  & +(3\tau_{1}^{z}\tau_{2}^{z}-\bm\tau_{1}\cdot\bm\tau_{2})\,\bm\sigma_{-}\cdot\bm\nabla(\bar{G}_{\pi}^{(2)}\,\mathcal{Y}_{\pi}(r)-\bar{G}_{\rho}^{(2)}\,\mathcal{Y}_{\rho}(r))\bigg\}\,,\label{eq:HPVTV-ex}\end{eqnarray}
 where $\bar{G}_{\ssst{X}}^{(i)}$ is defined as the product of a
strong coupling constant $g_{\ssst{XNN}}$ and its associated $\PVTV$
one $\bar{g}_{\ssst{X}}^{(i)}$;%
\footnote{In Ref.~\cite{Gudkov:1993yc}, the $\PVTV$ $\eta N\! N$ coupling
only contains an isoscalar part, so it does not contribute to the
isovector $\HPVTV$. However, this isovector piece, which gives a
different linear combination from the pion contribution, is needed
in order to render the $\tau_{+}^{z}\,\bm\sigma_{-}$ and $\tau_{-}^{z}\,\bm\sigma_{+}$
operators independent.%
} for instance, $\bar{G}_{\pi}^{(0)}=\gpiNN\,\bar{g}_{\pi}^{(0)}$.

One sees that the general operator structure in Eq. (\ref{eq:HPVTV-contact}),
based only on symmetry considerations, is fully reproduced by the
one-meson exchange scheme containing the lowest-lying pseudoscalar
and vector mesons in both isovector ($\pi$ and $\rho$) and isoscalar
($\eta$ and $\omega$) sectors. The ten coupling constants in Eq.~(\ref{eq:HPVTV-contact})
find their counterparts in the ten $\PVTV$ meson-nucleon coupling
constants. Eq.~(\ref{eq:HPVTV-ex}) has the advantage that it not
only has the most general operator structure, but it also provides
a link to the meson-exchange picture which provides some insight.
We finally note that one-kaon exchange does not contribute to the
strangeness-conserving $N\! N$ interaction.

\section{Deuteron EDM}

Because the $\PVTV$ interaction induces a small $P$-wave admixture
to the deuteron wave function, it leads to a nonvanishing matrix element
of the charge dipole operator. In addition, since the proton and the
neutron also have an EDM, a disentanglement of one- and two-body contributions,
\begin{equation}
d_{\mathcal{D}}=d_{\mathcal{D}}^{(1)}+d_{\mathcal{D}}^{(2)}\,\,,\end{equation}
 is necessary to make contact to the underlying $\PVTV$ physics.
In the following, we shall use the $\PVTV$ $N\! N$ interaction $\HPVTV$
constructed in the previous section to calculate $d_{\mathcal{D}}^{(2)}$.
We will use the same meson-exchange picture as a guideline to give
an estimate of $d_{\mathcal{D}}^{(1)}$. The final result for $d_{\mathcal{D}}$
can then be expressed in terms of the $\PVTV$ meson-nucleon coupling
constants. EDMs are expressed in units of $e$--fm for the remainder
of the paper.

\subsection{Two-Body Contributions}

For the two-body part, the dominant contribution comes from the polarization
effect: In leading order in the perturbation, it is the matrix element
of the charge dipole operator evaluated between the unperturbed deuteron
state $\ket{\mathcal{D}}$ (mainly $^{3}S_{1}$-wave with some $6\%$
$^{3}D_{1}$-wave) and the admixed $P$-wave component $\ket{\widetilde{{\mathcal{D}}}}$,
\textit{viz.}\begin{equation}
d_{\mathcal{D}}^{(pol)}=\sqrt{\frac{1}{6}}\,\bra{\mathcal{D}}|\tau_{-}^{z}\, e\,\bm r|\ket{\widetilde{\mathcal{D}}}\,,\label{eq:EDM-pol}\end{equation}
 where $\bm r=\bm r_{1}-\bm r_{2}$ and {}``||'' denotes the reduced
matrix element. Because the charge dipole operator conserves the total
spin, $\ket{\widetilde{{\mathcal{D}}}}$ has to be the $^{3}P_{1}$
state. The isospin and spin selection rules then dictate that only
the operator $\tau_{-}^{z}\,\bm\sigma_{+}$ in $\HPVTV$ can induce
such an admixture to the deuteron.

In order to examine the model dependence of the matrix element, the
numerical calculation is performed with three high-quality local potential
models: Argonne $v_{18}$ (A$v_{18}$)~\cite{Wiringa:1995wb}, and
the Nijmegen models Reid93 and Nijm II~\cite{Stoks:1994wp}. The
results \begin{subequations}
\begin{eqnarray}
   d_{\mathcal{D}}^{(pol)} & = & 1.43\times10^{-2}\,\bar{G}_{\pi}^{(1)}
                                +1.59\times10^{-3}\,\bar{G}_{\eta}^{(1)}
                                +6.25\times10^{-4}\,\bar{G}_{\rho}^{(1)}
                                -5.96\times10^{-4}\,\bar{G}_{\omega}^{(1)}\,,
\label{eq:Av18}\\
                           & = & 1.45\times10^{-2}\,\bar{G}_{\pi}^{(1)}
                                +1.68\times10^{-3}\,\bar{G}_{\eta}^{(1)}
                                +6.83\times10^{-4}\,\bar{G}_{\rho}^{(1)}
                                -6.53\times10^{-4}\,\bar{G}_{\omega}^{(1)}\,,
\label{eq:Reid93}\\
                           & = & 1.47\times10^{-2}\,\bar{G}_{\pi}^{(1)}
                                +1.72\times10^{-3}\,\bar{G}_{\eta}^{(1)}
                                +7.50\times10^{-4}\,\bar{G}_{\rho}^{(1)}
                                -7.19\times10^{-4}\,\bar{G}_{\omega}^{(1)}\,,
\label{eq:Nijm II}
\end{eqnarray}
\end{subequations} for A$v_{18}$, Reid93, and Nijm II, respectively, show a relatively
model-independent pattern. Judging from the coefficients for the different
mesons, pion exchange dominates the result. The much smaller sensitivity
of $d_{\mathcal{D}}^{(2)}\simeq d_{\mathcal{D}}^{(pol)}$ to heavy-meson
exchanges guarantees that pion-exchange is a good approximation here
(this may not be true for $d_{\mathcal{D}}^{(1)}$, a point we address
below).

\begin{figure}
\includegraphics[%
  scale=0.4]{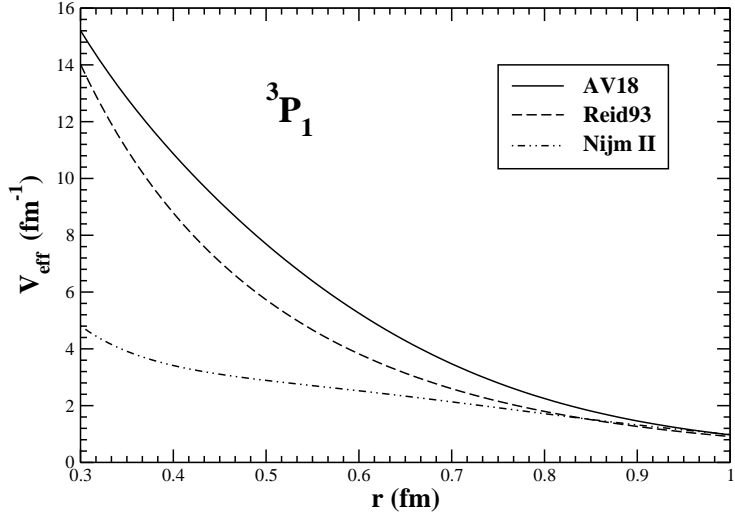}

\caption{The comparison of three different effective strong potentials in
the $^{3}P_{1}$ channel.}

\label{cap:Veff's}
\end{figure}

The slight difference in the results for these models can be attributed
to their softness at the intermediate range where the deuteron wave
function (which agrees well for these potential models) has most of
the overlap with the Yukawa functions. Fig.~\ref{cap:Veff's} compares
the effective potential $V_{eff}(r)=V_{\ssst{\mathrm{{S}}}}(r)+L(L+1)/(\mN r^{2})$
of these models in the $^{3}P_{1}$ channel ($L=1$), which determines
the radial behavior of the $\PVTV$ admixture in the inhomogeneous
Schr\"{o}dinger equation \begin{equation}
(T+V_{eff}^{(3P1)})\ket{\widetilde{\mathcal{D}}}=\HPVTV\ket{\mathcal{D}}\,.\end{equation}
 Among these three models, Nijm II is the softest one within the range
of about $0.3\sim1.0$~fm, so it gives the largest result, while
A$v_{18}$, the hardest one, gives the smallest result. As the heavy-meson
exchange is very sensitive to the wave function at short range, its
model dependence is more apparent compared to the pion-exchange case.
Our result for the coefficient of $\bar{G}_{\pi}^{1}$ is consistent
with two earlier predictions: 0.010--0.026 obtained by Avishai~\cite{Avishai:1985},
who used strong potential models from before the 70s, and $0.019$
obtained by Khriplovich and Korkin~\cite{Khriplovich:1999qr}, who
assumed the zero-range approximation for the deuteron and a free $^{3}P_{1}$
wave function. Their number can be considered as an upper bound.

The meson-exchange effects, in the form of two-body exchange charges,
give contributions of the form \begin{equation}
d_{\mathcal{D}}^{(ex)}=\sqrt{\frac{1}{6}}\,\left(\bra{\mathcal{D}}|\int d^{3}x\,\rho^{(2)}(x)\,\bm x|\ket{\widetilde{\mathcal{D}}}+\bra{\mathcal{D}}|\int d^{3}x\,\tilde{\rho}^{(2)}(x)\,\bm x|\ket{\mathcal{D}}\right)\,,\label{eq:EDM-ex}\end{equation}
 where the first term corresponds to adding the normal ($P$- and
$T$-even) exchange charge $\rho^{(2)}$ to Eq.~(\ref{eq:EDM-pol}),
and the $\PVTV$ exchange $\tilde{\rho}^{(2)}$, induced by $\HPVTV$,
is included via the second term. Compared with the one-body charge,
which is $\mathcal{O}(1)$, $\rho^{(2)}$ can be ignored since it
gives a correction of $\mathcal{O}(1/\mN)^{3}$ (see \textit{e.g.}
Ref.~\cite{Riska:1989bh}). On the other hand, since $\tilde{\rho}^{(2)}$
can be as large as $\mathcal{O}(1/\mN^{2})$, and its contribution
is evaluated within unperturbed deuteron wave functions, its significance
should be investigated.

As indicated by the dominance of pion exchange observed above, and
also in view of the suppression of heavy-meson exchange currents found
in the study of the ($P$-odd, $T$-even) deuteron anapole moment~\cite{Liu:2003au},
the consideration of the pion sector is sufficient for the two-body
exchange effects. Attaching a photon to every possible line in the
one-pion exchange diagram which leads to $\HPVTV^{(\pi)}$, the exchange
charge can then to $\mathcal{O}(1/\mN^{2})$ be identified, in configuration
space, as \begin{eqnarray}
\tilde{\rho}_{pair}^{(\pi)}(\bm x;\bm r_{1},\bm r_{2}) & = & \frac{e}{4\,\mN^{2}}\Big((1+\kappaS)\,\big(\bar{G}_{\pi}^{(0)}\,\bm\tau_{1}\cdot\bm\tau_{2}+\bar{G}_{\pi}^{(1)}\,\tau_{1}^{z}+\bar{G}_{\pi}^{(2)}\,(3\,\tau_{1}^{z}\,\tau_{2}^{z}-\bm\tau_{1}\cdot\bm\tau_{2})\big)\nonumber \\
 &  & +(1+\kappaV)\,(\bar{G}_{\pi}^{(0)}\,\tau_{2}^{z}+\bar{G}_{\pi}^{(1)}+2\,\bar{G}_{\pi}^{(2)}\,\tau_{2}^{z})\Big)\,\bm\sigma_{1}\cdot\bm\nabla_{x}\,\delta^{(3)}(\bm x-\bm r_{1})\,\mathcal{Y}_{\pi}(r)\nonumber \\
 &  & +(1\leftrightarrow2)\,,\\
\tilde{\rho}_{mesonic}^{(\pi)}(\bm x;\bm r_{1},\bm r_{2}) & = & -\frac{e}{4\,\mN^{2}}\, i\,(\bm\tau_{1}\times\bm\tau_{2})^{z}\,(\bar{G}_{\pi}^{(0)}-\bar{G}_{\pi}^{(2)})\,(\bm\sigma_{1}\cdot\bm\nabla_{1}+\bm\sigma_{2}\cdot\bm\nabla_{2})\nonumber \\
 &  & \times\big[\nabla_{1}^{2}-\nabla_{2}^{2}\,,\,\mathcal{Y}_{\pi}(r_{x1})\,\mathcal{Y}_{\pi}(r_{x2})\big]\,,\end{eqnarray}
 where the pair term refers to the diagram in which the photon couples
to an intermediate nucleon-antinucleon pair and the mesonic term refers
to the diagram in which the photon couples to the meson in flight;
$r=|\bm r_{1}-\bm r_{2}|$, $r_{x1(2)}=|\bm x-\bm r_{1(2)}|$. Numerically,
the contribution of these diagrams to the deuteron EDM is found to
be \begin{equation}
d_{\mathcal{D}}^{(ex)}\simeq9.40\times10^{-4}\,\bar{G}_{\pi}^{(1)}-5.28\times10^{-4}\,\bar{G}_{\pi}^{(0)}\,.\label{eq:pion-ex}\end{equation}
 Compared with $d_{\mathcal{D}}^{(pol)}$, this constitutes only a
few-percent correction.

Combining the results for $d_{\mathcal{D}}^{(pol)}$ and $d_{\mathcal{D}}^{(ex)}$,
we obtain for the two-body contribution to the deuteron EDM, in terms
of $\PVTV$ couplings, \begin{equation}
d_{\mathcal{D}}^{(2)}=d_{\mathcal{D}}^{(pol)}+d_{\mathcal{D}}^{(ex)}\simeq0.20\,\bar{g}_{\pi}^{(1)}+\mathcal{O}(\bar{g}_{\pi}^{(0)},\,\bar{g}_{\eta,\,\rho,\,\omega}^{(1)})\,,\label{eq:d-EDM(2)}\end{equation}
 with an error estimated as less than $5\%$.

\begin{figure}
\includegraphics[%
  scale=0.7]{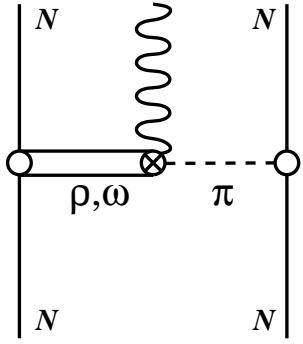}

\caption{The two-body contribution to $d_{\mathcal{D}}$ arising from the
first-order $\PVTV$ $\rho$- and $\omega$-$\pi\gamma$ couplings.
\label{cap:mesonic-PVTV}}
\end{figure}

Besides the usual exchange effects in which one of the meson-nucleon
couplings is $P$- and $T$-odd, another class of diagrams involving
a $\PVTV$ photon coupling to the exchanged mesons can also contribute.
Since pseudoscalar mesons cannot have such a $\PVTV$ coupling to
photons, the candidates in our current framework are $\PVTV$ $\rho\pi\gamma$,
$\omega\pi\gamma$, and $\rho\rho\gamma$ vertices. Assuming these
$\PVTV$ couplings are of the same order of magnitude, one can expect
a smaller contribution from the $\rho\rho\gamma$ vertex, because
the $\rho$ meson is much more massive and has a smaller strong coupling
to nucleons than the pion. Therefore, in order to estimate the size
of this type of contributions we evaluate the diagrams based on $\PVTV$
$\rho\pi\gamma$ and $\omega\pi\gamma$ vertices shown in Fig.~\ref{cap:mesonic-PVTV}.

Expressing the $\PVTV$ $\rho\pi\gamma$ and $\omega\pi\gamma$ Lagrangian
densities as \begin{eqnarray}
\mathcal{L}_{\ssst{\PVTV}}^{(\rho\pi\gamma)} & = & \frac{e\,\bar{g}_{\rho\pi\gamma}}{2m_{\rho}}\, F^{\alpha\beta}\,\bm\rho_{\alpha}\cdot\partial_{\beta}\bm\pi\,,\\
\mathcal{L}_{\ssst{\PVTV}}^{(\omega\pi\gamma)} & = & \frac{e\,\bar{g}_{\omega\pi\gamma}}{2m_{\omega}}\, F^{\alpha\beta}\,\omega_{\alpha}\,\partial_{\beta}\pi^{0}\,,\end{eqnarray}
 where two new $\PVTV$ coupling constants $\bar{g}_{\rho\pi\gamma}$
and $\bar{g}_{\omega\pi\gamma}$ are introduced, the associated exchange
charges are, in configuration space, \begin{eqnarray}
\tilde{\rho}_{mesonic}^{(\rho\pi\gamma')}(\bm x;\bm r_{1},\bm r_{2}) & = & \frac{e\grhoNN\gpiNN\bar{g}_{\rho\pi\gamma}}{4m_{\rho}\mN}\,\bm\tau_{1}\cdot\bm\tau_{2}\,(\bm\nabla\cdot\bm\nabla_{2})(\bm\sigma_{2}\cdot\bm\nabla_{2})\,\mathcal{Y}_{\rho}(r_{x1})\mathcal{Y}_{\pi}(r_{x2})+(1\leftrightarrow2)\,,\\
\tilde{\rho}_{mesonic}^{(\omega\pi\gamma')}(\bm x;\bm r_{1},\bm r_{2}) & = & \frac{e\gomegaNN\gpiNN\bar{g}_{\omega\pi\gamma}}{4m_{\omega}\mN}\,(\bm\nabla\cdot\bm\nabla_{2})(\bm\sigma_{2}\cdot\bm\nabla_{2})\,\mathcal{Y}_{\omega}(r_{x1})\mathcal{Y}_{\pi}(r_{x2})+(1\leftrightarrow2)\,.\end{eqnarray}
 The numerical calculation, using the A$v_{18}$ potential, gives
an EDM contribution of about $2.3\times10^{-3}\,(\bar{g}_{\rho\pi\gamma}-\bar{g}_{\omega\pi\gamma})$.
Since the coefficient is two orders of magnitude smaller than the
leading coefficient of $\bar{g}_{\pi}^{(1)}$ in Eq.~(\ref{eq:d-EDM(2)}),
we shall ignore these mesonic $\PVTV$ effects for the rest of this
work.

\subsection{One-Body Contributions}

The total one-body contribution to the deuteron EDM is simply the
sum of the proton and neutron EDMs, \textit{i.e.}\begin{equation}
d_{\mathcal{D}}^{(1)}=d_{p}+d_{n}\,.\end{equation}
 Our goal in this section is to evaluate $d_{p}$ and $d_{n}$ in
a manner consistent with the framework used for the $\PVTV$ $N\! N$
interaction.

The nucleon EDM has a wide variety of sources such as the QCD $\bar{\theta}$
term, quark EDMs and chromo-EDMs (CEDMs), Weinberg three-gluon operator,
and four-quark contact interactions, therefore, its evaluation requires
good knowledge of nonperturbative dynamics of confined quarks, which
is still not available. A commonly used method of estimate is to evaluate
the hadronic loop diagrams, in which meson and baryon degrees of freedom
are used to describe the dynamics and the dependence on the $\PVTV$
mechanisms at the quark-gluon level is subsumed in the $\PVTV$ meson-nucleon
coupling constants. This approach has been applied extensively to
the neutron EDM in various contexts (see, for example, Refs.~\cite{Barton:1969gi,Crewther:1979pi,McKellar:1987tf,He:1989xj,Valencia:1990cm,Pich:1991fq,Khatsymovsky:1992yg,Khriplovich:1996gk,Borasoy:2000pq}).
Here we apply it to both the proton and the neutron EDM, with the
inclusion of vector mesons.

\begin{figure}
\begin{tabular}{ccc}
\includegraphics[%
  scale=0.7]{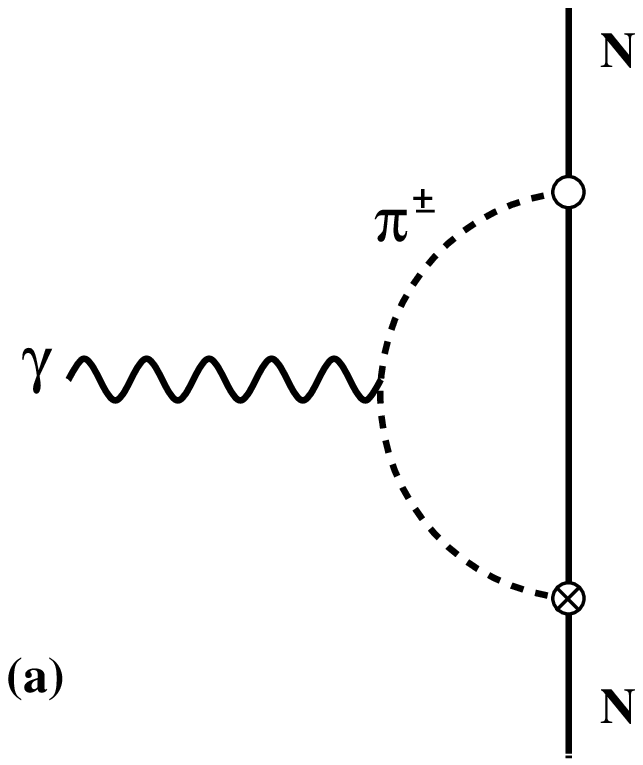}&
\includegraphics[%
  scale=0.7]{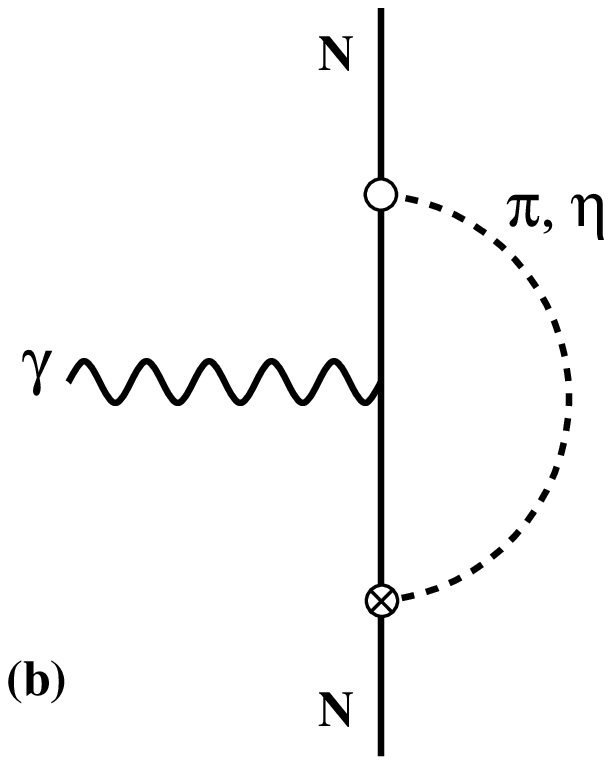}&
\includegraphics[%
  scale=0.7]{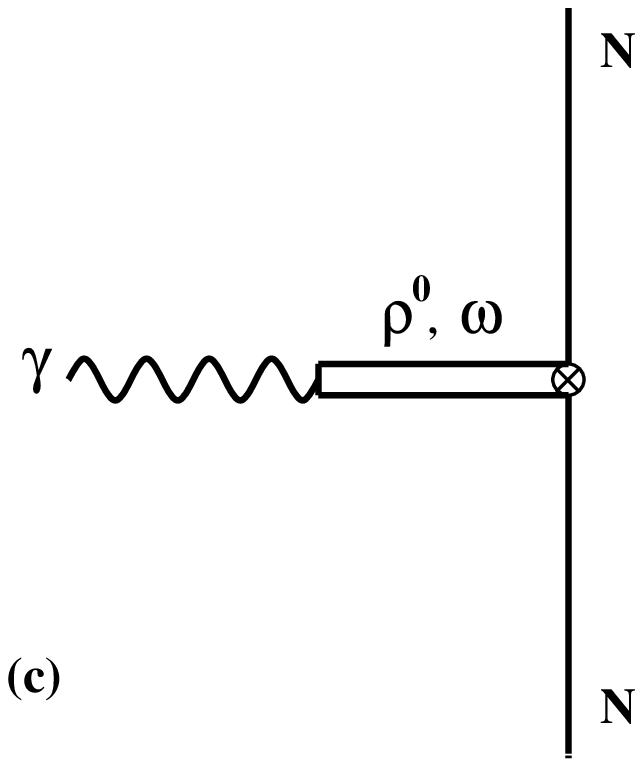}\tabularnewline
\end{tabular}

\caption{Hadronic loop diagrams which contribute to the nucleon EDM. \label{cap:loop diagrams}}
\end{figure}

The loop diagrams containing a virtual pseudoscalar meson are classified
as in Fig.~\ref{cap:loop diagrams}(a) and Fig.~\ref{cap:loop diagrams}(b),
where the photon couples to the charged pseudoscalar meson in the
former and to the intermediate nucleon in the latter case. Defining
the hadronic loop contribution to the nucleon EDM as \begin{equation}
d_{\ssst{N}}^{(had)}\equiv\frac{e}{4\pi^{2}\mN}\left(\delta_{\ssst{S}}\,\frac{1}{2}+\delta_{\ssst{V}}\,\frac{\tau^{z}}{2}\right)\,,\label{eq:dN-hadronic}\end{equation}
 the results for the corresponding diagrams are%
\footnote{Kaon loops can also contribute~\cite{McKellar:1987tf,Valencia:1990cm,He:1989xj,Pich:1991fq,Borasoy:2000pq},
and be easily added to our results.%
}\begin{eqnarray}
\delta_{\ssst{S}}^{(a)} & = & 0\,,\label{eq:(a)-S}\\
\delta_{\ssst{V}}^{(a)} & = & -2(\bar{G}_{\pi}^{(0)}-\bar{G}_{\pi}^{(2)})\mathcal{I}_{0}^{(\pi)}\,,\label{eq:(a)-V}\\
\delta_{\ssst{S}}^{(b)} & = & (3\bar{G}_{\pi}^{(0)}+\bar{G}_{\pi}^{(1)})\mathcal{I}_{1}^{(\pi)}-(3\kappaS\bar{G}_{\pi}^{(0)}+\kappaV\bar{G}_{\pi}^{(1)})\mathcal{I}_{2}^{(\pi)}\nonumber \\
 &  & +(\bar{G}_{\eta}^{(0)}+\bar{G}_{\eta}^{(1)})\mathcal{I}_{1}^{(\eta)}-(\kappaS\bar{G}_{\eta}^{(0)}+\kappaV\bar{G}_{\eta}^{(1)})\mathcal{I}_{2}^{(\eta)}\,,\label{eq:(b)-S}\\
\delta_{\ssst{V}}^{(b)} & = & (-\bar{G}_{\pi}^{(0)}+\bar{G}_{\pi}^{(1)}+4\bar{G}_{\pi}^{(2)})\mathcal{I}_{1}^{(\pi)}-(-\kappaV\bar{G}_{\pi}^{(0)}+\kappaS\bar{G}_{\pi}^{(1)}+4\kappaV\bar{G}_{\pi}^{(2)})\mathcal{I}_{2}^{(\pi)}\nonumber \\
 &  & +(\bar{G}_{\eta}^{(0)}+\bar{G}_{\eta}^{(1)})\mathcal{I}_{1}^{(\eta)}-(\kappaV\bar{G}_{\eta}^{(0)}+\kappaS\bar{G}_{\eta}^{(1)})\mathcal{I}_{2}^{(\eta)}\,.\label{eq:(b)-V}\end{eqnarray}
 The three distinct loop integrals involving an $i$-type pseudoscalar
meson, $\mathcal{I}_{0}^{(i)}$, $\mathcal{I}_{1}^{(i)}$, and $\mathcal{I}_{2}^{(i)}$,
correspond to the cases where the photon couples to the pseudoscalar
meson, the nucleon Dirac, and the nucleon Pauli form factor, respectively.
They are evaluated as \begin{eqnarray}
\mathcal{I}_{0}^{(i)} & = & -1-(1-x_{i}^{2})\,\ln x_{i}+x_{i}^{2}\,(3-x_{i}^{2})\,\mathcal{F}(x_{i}^{2})\nonumber \\
 & \xrightarrow[x_{i}\ll1]{} & -\ln x_{i}-1+\frac{3\,\pi}{4}\, x_{i}+x_{i}^{2}\,\ln x_{i}+\mathcal{O}(x_{i}^{2})\,,\label{eq:loop-(a)}\\
\mathcal{I}_{1}^{(i)} & = & \frac{1}{2}-\frac{1}{2}\, x_{i}^{2}\,\ln x_{i}-x_{i}^{2}\,(1-\frac{1}{2}x_{i}^{2})\,\mathcal{F}(x_{i}^{2})\nonumber \\
 & \xrightarrow[x_{i}\ll1]{} & \frac{1}{2}-\frac{\pi}{4}\, x_{i}-\frac{1}{2}\, x_{i}^{2}\,\ln x_{i}+\mathcal{O}(x_{i}^{2})\,,\label{eq:loop-(b1)}\\
\mathcal{I}_{2}^{(i)} & = & \frac{11}{16}+\frac{3}{8}\, x_{i}^{2}-\frac{1}{16}\, x_{i}^{2}\,(1+3\, x_{i}^{2})\,\ln x_{i}-x_{i}^{2}\,(1+\frac{5}{8}\, x_{i}^{2}+\frac{3}{8}\, x_{i}^{2})\,\mathcal{F}(x_{i}^{2})\nonumber \\
 & \xrightarrow[x_{i}\ll1]{} & \frac{11}{16}-\frac{\pi}{4}\, x_{i}-\frac{1}{8}x_{i}^{2}\,\ln x_{i}+\mathcal{O}(x_{i}^{2})\,,\label{eq:loop-(b2)}\\
\mathcal{F}(s) & = & \frac{1}{\sqrt{4\, s+s^{2}}}\left(\tan^{-1}\left[\frac{2-s}{\sqrt{4\, s+s^{2}}}\right]+\tan^{-1}\left[\frac{s}{\sqrt{4\, s+s^{2}}}\right]\right)\,,\end{eqnarray}
 where $x_{i}=m_{i}/\mN$. Since both meson and nucleon form factors
were taken to be constant off the mass shell, and since form factors
fall off as the square of the four-momentum transfer increases, these
results should be viewed as an upper bound~\cite{McKellar:1987tf}.

From Eqs.~(\ref{eq:loop-(a)})--(\ref{eq:loop-(b2)}), one observes
that only $\mathcal{I}_{0}^{(\pi)}$ has a non-analytic term, \textit{i.e.}
$\ln x_{\pi}$, in the chiral limit, $m_{\pi}\rightarrow0$. The mathematical
reason is that Fig.~\ref{cap:loop diagrams}(a) contains more pion
propagators than Fig.~\ref{cap:loop diagrams}(b), which is responsible
for the infrared divergence in the soft-pion limit~\cite{Crewther:1979pi}.
Therefore, the contribution to the nucleon EDM involving chiral logarithms
is purely isovector, \begin{equation}
d_{\ssst{N}}^{(\pi)}=-\frac{e\tau^{z}}{4\pi^{2}\mN}\,(\bar{G}_{\pi}^{(0)}-\bar{G}_{\pi}^{(2)})\ln\left(\frac{m_{N}}{m_{\pi}}\right)\,.\end{equation}
 This implies that the deuteron EDM receives no one-body contribution
from loop diagrams involving $\pi$ and $\eta$ mesons in the chiral
limit. Furthermore, when the neutron EDM is considered, the constant
terms in $\mathcal{I}_{0}$ and $\mathcal{I}_{1}$ exactly cancel,
as has been pointed out in Ref.~\cite{Pich:1991fq}. However, this
is not true for the proton.

Because chiral symmetry is explicitly broken by the pion mass, it
is interesting to compare the chiral logarithm with other, analytic,
terms when realistic parameters are used. For example, taking $x_{\pi}=140/940$
in Eq.~(\ref{eq:loop-(a)}), one gets $\mathcal{I}_{0}^{(\pi)}\simeq1.19$,
which is about $40\%$ smaller than $-\ln(m_{\pi}/\mN)\simeq1.90$.
This sizable difference typically sets the scale for the theoretical
uncertainty. The same conclusion can be drawn from the work by Barton
and White~\cite{Barton:1969gi} who, motivated by the success of
sideways dispersion relations for the nucleon Pauli form factors~\cite{Drell:1965hg},
applied the same technique with the same parameters to the neutron
EDM problem. This analysis, involving mainly the threshold pion-photoproduction
amplitude, is actually similar to the evaluation of type (a) loop
diagram with soft pions, and in fact produces the same chiral logarithm.
Compared with the leading term which gave $d_{n}\simeq0.8\times10^{-11}$
$e$--fm (a value, in our notation, $\bar{g}_{\pi}^{(0)}-\bar{g}_{\pi}^{(2)}=1.2\times10^{-10}$
was used as input), their full analysis predicted $d_{n}\simeq0.5\times10^{-11}$
$e$--fm, which is also some $40\%$ smaller. While this may be just
an accident, it does signal a potentially large theoretical uncertainty
for the nucleon EDM.

In order to estimate the relevance of the vector-meson degrees of
freedom to the nucleon EDM, we consider the diagrams illustrated in
Fig.~\ref{cap:loop diagrams}(c). These contributions can be roughly
estimated by the assumption of VMD, which leads to a dispersion-theory
analysis of the $\rho^{0}$ and $\omega$ poles in the time-like region.
The deuteron is only sensitive to the isoscalar sector, for which,
in the case of the nucleon Pauli form factor, the naive VMD model
works rather well. The vector-meson contributions to the isovector
nucleon EDM should, however, also be added as a correction to the
leading result from the pion-loop calculation, which is equivalent
to including the $2\pi$ continuum in the dispersion-theory analysis.

The required vector-meson-photon conversion mechanism is introduced
by the Lagrangian density \begin{equation}
\mathcal{L}_{\ssst{\mathrm{VMD}}}=\frac{e}{2f_{\rho}}F^{\mu\nu}F_{\mu\nu}^{(\rho)}+\frac{e}{2f_{\omega}}F^{\mu\nu}F_{\mu\nu}^{(\omega)}\,,\end{equation}
 where the $F^{\mu\nu}$'s denote the field tensors for the photon
and the $\rho^{0}$ and $\omega$ mesons; the constants $f_{\rho}=5.00$
and $f_{\omega}=17.05$ are determined from the decay widths $\Gamma_{\rho,\omega\rightarrow e^{+}e^{-}}=6.85,0.60$
MeV~\cite{Hagiwara:2002fs} by $\Gamma_{x\rightarrow e^{+}e^{-}}=4\pi\alpha^{2}\mN/(3f_{x}^{2})$.
Then the vector-meson contributions to the nucleon EDM are evaluated
as \begin{eqnarray}
\delta_{\ssst{S}}^{(c)} & = & \frac{4\pi^{2}}{f_{\rho}\grhoNN}\bar{G}_{\rho}^{(1)}+\frac{4\pi^{2}}{f_{\omega}\gomegaNN}\bar{G}_{\omega}^{(0)}\,,\label{eq:(c)-S}\\
\delta_{\ssst{V}}^{(c)} & = & \frac{4\pi^{2}}{f_{\rho}\grhoNN}(\bar{G}_{\rho}^{(0)}+2\bar{G}_{\rho}^{2})+\frac{4\pi^{2}}{f_{\omega}\gomegaNN}\bar{G}_{\omega}^{(1)}\,.\label{eq:(c)-V}\end{eqnarray}

Keeping in mind the caveat of a possibly large theoretical error,
we nevertheless take a more adventurous point of view and include
also the analytic terms in $\mathcal{I}_{0}$ and $\mathcal{I}_{1}$,
\textit{i.e.} $\mathcal{I}_{0}^{(\pi)}\simeq1.19$, $\mathcal{I}_{1}^{(\pi)}\simeq0.41$,
and $\mathcal{I}_{1}^{(\eta)}\simeq0.28$, but we neglect the part
from the nucleon Pauli form factor, so we set $\mathcal{I}_{2}^{(\pi)}=\mathcal{I}_{2}^{(\eta)}=0$.
Collecting the results from Eqs.~(\ref{eq:dN-hadronic}), (\ref{eq:(a)-S})--(\ref{eq:(b)-V}),
(\ref{eq:(c)-S})--(\ref{eq:(c)-V}), the total one-body contribution
to the deuteron EDM is evaluated as \begin{eqnarray}
d_{\mathcal{D}}^{(1)} & = & 2.18\times10^{-3}\,(3\,\bar{G}_{\pi}^{(0)}+\bar{G}_{\pi}^{(1)})+1.49\times10^{-3}\,(\bar{G}_{\eta}^{(0)}+\bar{G}_{\eta}^{(1)})\nonumber \\
 &  & +1.53\times10^{-2}\,\bar{G}_{\rho}^{(1)}+1.49\times10^{-3}\,\bar{G}_{\omega}^{(0)}\,.\label{eq:many-qaurk}\end{eqnarray}
 In terms of the $\PVTV$ meson-nucleon coupling constants, the result
is \begin{equation}
d_{\mathcal{D}}^{(1)}=0.03\,\bar{g}_{\pi}^{(1)}+0.09\,\bar{g}_{\pi}^{(0)}+0.04\,\bar{g}_{\rho}^{(1)}+0.01\,\bar{g}_{\omega}^{(0)}+\mathcal{O}(\bar{g}_{\eta}^{(0,1)})\,.\label{eq:d-EDM(1)}\end{equation}

\section{Discussion}

Combining Eqs.~(\ref{eq:d-EDM(2)}) and (\ref{eq:d-EDM(1)}), we
arrive at our final estimate for the deuteron EDM: \begin{equation}
d_{\mathcal{D}}=(0.20+0.03)\,\bar{g}_{\pi}^{(1)}+0.09\,\bar{g}_{\pi}^{(0)}+0.04\,\bar{g}_{\rho}^{(1)}+0.01\,\bar{g}_{\omega}^{(0)}\,.\label{eq:d-D-final}\end{equation}
 while our results at the same time imply the following predictions
for the proton and neutron EDMs: \begin{eqnarray}
d_{p} & = & -0.08(\bar{g}_{\pi}^{(0)}-\bar{g}_{\pi}^{(2)})+0.03(\bar{g}_{\pi}^{(0)}+\bar{g}_{\pi}^{(1)}+2\bar{g}_{\pi}^{(2)})+3\times10^{-3}(\bar{g}_{\eta}^{(0)}+\bar{g}_{\eta}^{(1)})\\
 &  & +0.02(\bar{g}_{\rho}^{(0)}+\bar{g}_{\rho}^{(1)}+2\bar{g}_{\rho}^{(2)})+6\times10^{-3}(\bar{g}_{\omega}^{(0)}+\bar{g}_{\omega}^{(1)})\,,\nonumber \\
d_{n} & = & 0.14(\bar{g}_{\pi}^{(0)}-\bar{g}_{\pi}^{(2)})-0.02(\bar{g}_{\rho}^{(0)}-\bar{g}_{\rho}^{(1)}+2\bar{g}_{\rho}^{(2)})+6\times10^{-3}(\bar{g}_{\omega}^{(0)}-\bar{g}_{\omega}^{(1)})\,.\label{eq:d-n-final}\end{eqnarray}

The leading contribution to $d_{\mathcal{D}}$, $0.20\,\bar{g}_{\pi}^{(1)}$,
due to the $\PVTV$ $N\! N$ interaction, including the exchange charges
and calculated by using state-of-the-art wave functions, is $25\%$
smaller than the result assuming the zero-range approximation~\cite{Khriplovich:1999qr}
which was adopted for an analysis on $C\! P$ violation models in
Ref.~\cite{Lebedev:2004va}. The remaining contributions come from
the proton and neutron EDMs. These terms have a sizable theoretical
uncertainty, which could be as large as $40\%$. The non-vanishing
dependence on $\bar{g}_{\pi}^{(0)}$, which arises from including
analytic terms in the hadronic loop calculations, sets the stage for
the QCD $\bar{\theta}$ term to play a role in the deuteron. Using
the prediction by Crewther~\emph{et al.}~\cite{Crewther:1979pi}
that $\bar{g}_{\pi}^{(0)}\simeq0.027\bar{\theta}$, one gets a $d_{\mathcal{D}}\sim2\times10^{-3}\bar{\theta}$
dependence on $\bar{\theta}$, which is about three times larger than
the QCD sum-rule calculation \cite{Pospelov:1999ha,Lebedev:2004va}.
The dependence on the vector-meson couplings, though suppressed at
the two-body level, enter the final result through the nucleon EDM
where it could be sizable. An important issue in this respect is the
size of the $\PVTV$ $\rho$, $\omega$ (and $\eta$ for that matter)
coupling constants compared to those of the pion. An argument by Gudkov~\emph{et
al.} suggested that these vector-meson coupling constants are less
significant \cite{Gudkov:1993yc}, while a recent work by Pospelov
based on QCD sum rules gave the {}``best'' values for $\bar{g}_{\rho,\omega}$
of the same order of magnitude as $\bar{g}_{\pi}$~\cite{Pospelov:2001ys};
and these two works surprisingly have opposite predictions about the
relative importance of $\bar{g}_{\pi^{\pm}}/\bar{g}_{\pi^{0}}$. Furthermore,
work on the $P$-odd, $T$-even $N\! N$ interaction implied vector
couplings at least equally important and preferably larger than their
pseudoscalar counterparts (see \textit{e.g.} Refs.~\cite{Desplanques:1980hn,Haxton:2001mi,Haxton:2001zq}).
Therefore, until consensus is reached, these vector-meson contributions
should still be kept for maintaining a greater generality.

In order to connect expression Eq.~(\ref{eq:d-D-final}) for the
deuteron EDM to the underlying $C\! P$ violation, the $\PVTV$ meson-nucleon
coupling constants have to be expressed in terms of parameters at
particle-physics level, such as the QCD $\bar{\theta}$ term, quark
EDMs and CEDMs, \textit{etc.} These quantities have a plethora of
predictions from extensions of the Standard Model. Because all the
EDM measurements to date only resulted in upper bounds, it is a popular
practice to use these experimental limits to derive the corresponding
bounds for one particular source of $C\! P$ violation while turning
other possibilities off in an \emph{ad hoc} fashion. Even though this
simplification is legitimate to some extent, one might obtain an overconstraint
by excluding possible cancellations between various $C\! P$-violation
sources.

The deuteron and neutron results illustrate how limits on their EDMs
could be used to provide tight constraints on a specific model of
$C\! P$ violation, such as the one in Ref.~\cite{Pospelov:2001ys}.
For supersymmetric models in which the Pecci-Quinn symmetry is evoked
to remove the QCD $\bar{\theta}$ term, the quark CEDMs are the dominant
contributors to the $\PVTV$ meson-nucleon coupling constants, compared
to the three-gluon and four-quark operators. Therefore, all the $\bar{g}$'s
can be expressed in terms of the $d_{q}^{c}$'s. Using the {}``best''
values recommended in Ref.~\cite{Pospelov:2001ys}: $\bar{g}_{\pi}^{(1)}\simeq20\, d_{-}^{c}$,
$\bar{g}_{\pi}^{(0)}\simeq4\, d_{+}^{c}$, $\bar{g}_{\rho}^{(0)}\simeq13.3\, d_{+}^{c}$,
$\bar{g}_{\rho}^{(1)}\simeq-8.6\, d_{-}^{c}$, $\bar{g}_{\omega}^{(0)}\simeq-8.6\, d_{+}^{c}$,
$\bar{g}_{\omega}^{(1)}\simeq-13.3\, d_{-}^{c}$, where $d_{\pm}^{c}=d_{u}^{c}\pm d_{d}^{c}$,%
\footnote{In Ref. \cite{Pospelov:2001ys}, the $\PVTV$ vector-meson-nucleon
couplings are defined to have the same dimension as the EDM. The conversion
to our definition is a factor of $2\,\mN$~\cite{-best_values}.%
} the deuteron and neutron EDMs can be completely expressed in terms
of the CEDMs of the up and down quarks, \textit{viz.}\begin{eqnarray}
d_{\mathcal{D}} & = & -4.67\, d_{d}^{c}+5.22\, d_{u}^{c}\,,\\
d_{n} & = & -0.01\, d_{d}^{c}+0.49\, d_{u}^{c}\,.\end{eqnarray}
 Thus, these two EDM measurements probe different linear combinations
of $d_{d}^{c}$ and $d_{u}^{c}$ in this case. Moreover, the deuteron
could be significantly more sensitive than the neutron. This example
is clearly oversimplified, however, judging from the general expressions
Eqs.~(\ref{eq:d-D-final}) and (\ref{eq:d-n-final}), one expects
that, barring unnatural and accidental cancellations, the deuteron
is competitive to the neutron in sensitivity to $C\! P$ violation.
Furthermore, the deuteron EDM involves different $\PVTV$ coupling
constants, and hence in general will be complementary with respect
to the information about $C\! P$ violation that can be probed with
the neutron.

In conclusion, it should be realized that the theoretical uncertainties,
especially in the results for $d_{p}$ and $d_{n}$ and hence in the
one-body contribution to $d_{\mathcal{D}}$, are significant. The
calculation of an atomic or nuclear EDM involves a broad range of
physics, including the problematic strong interaction at the nuclear
and subnuclear scale. In this respect, it is relevant that efforts
have been renewed recently to attack the neutron EDM in lattice QCD~\cite{Guadagnoli:2002nm}.
In general, improved treatments of the hadronic physics, which can
bridge the phenomenology of the neutron EDM and $\PVTV$ nuclear forces
with the underlying particle physics, are of central interest.

\begin{acknowledgments}
We are grateful to A.E.L. Dieperink for helpful discussions, and to
our experimental colleagues, in particular C.J.G. Onderwater, K. Jungmann,
and H.W. Wilschut, for their interest and for comments. We also thank
P. Herczeg for raising the issue of $\PVTV$ photon couplings to mesons.
\end{acknowledgments}
\appendix

\section{Deuteron MQM}

Besides the EDM, the $\PVTV$ $N\! N$ interaction can also induce
$P$- and $T$-odd electromagnetic moments of higher multipolarity,
that is, $C3$, $C5$, and $M2$, $M4$, \textit{etc.} For a spin-1
object such as the deuteron, a nonzero $M2$ magnetic quadrupole moment
(MQM) is therefore another signature of $C\! P$ violation. Approximating
the nuclear electromagnetic current as purely one-body, \textit{i.e.}
ignoring the meson-exchange currents, the MQM operator can be expressed,
in a Cartesian basis, as \begin{equation}
M_{mn}=\frac{e}{2\mN}\left\{ \mu\left[3\, r_{m}\,\sigma_{n}+3\, r_{n}\,\sigma_{m}-2\,\bm\sigma\cdot\bm r\,\delta_{mn}\right]+2\, q\left(r_{m}\, L_{n}+r_{n}\, L_{m}\right)\right\} \,,\end{equation}
 where $\mu$, $q$, and $L$ denote the nucleon magnetic moment,
charge (in uints of $e$), and orbital angular momentum, respectively~\cite{Khriplovich:1999qr}.
The deuteron MQM, defined by \begin{equation}
M_{\mathcal{D}}=2\bra{\mathcal{D},J_{z}=1}\sum_{i=1}^{2}M_{zz}(i)\ket{\widetilde{{\mathcal{D}}},J_{z}=1}\,,\end{equation}
 can then be evaluated once the $^{1}P_{1}$ and $^{3}P_{1}$ parity
admixtures have been calculated. Assuming $\PVTV$ one-pion exchange
only, and using the A$v_{18}$ strong potential gives the numerical
result \begin{equation}
M_{\mathcal{D}}=0.051\,\muS\,\bar{g}_{\pi}^{(0)}+(0.031\,\muV+0.003)\,\bar{g}_{\pi}^{(1)}\,,\end{equation}
 in units of $e$--fm$^{2}$. The model-dependency is at the $1\%$
level, similar to the EDM calculation.

Although the isoscalar spin current leads to a rather large matrix
element (in the zero-range approximation of Ref.~\cite{Khriplovich:1999qr},
it is three times the isovector one), the isoscalar magnetic moment
renders the resulting $\bar{g}_{\pi}^{(0)}$ coefficient, 0.04, smaller
than the $\bar{g}_{\pi}^{(1)}$ coefficient, 0.15, which is dominated
by the isovector spin current from the large isovector magnetic moment.
The orbital motion adds only a small correction to the $\bar{g}_{\pi}^{(1)}$
term through the deuteron $D$-wave component.

While a sensitive experiment to measure $M_{\mathcal{D}}$ appears
as least as formidable as for $d_{\mathcal{D}}$, it might be contemplated
with deuterium atoms, because the MQM, unlike the EDM, is not screened
by the electron. An interesting theoretical point is that, since the
nucleon itself has no quadrupole moment, the deuteron MQM is a rather
clean probe of the $\PVTV$ $N\! N$ interaction, and in particular
of $\bar{g}_{\pi}^{(1)}$.

\bibliographystyle{apsrev}
\bibliography{/home/cpliu/research/papers/d_EDM/EDM,/home/cpliu/research/papers/d_EDM/CPV,/home/cpliu/research/papers/d_EDM/NN,/home/cpliu/research/papers/d_EDM/MN,/home/cpliu/research/papers/d_EDM/general,/home/cpliu/research/papers/d_EDM/AM,/home/cpliu/research/papers/d_EDM/MEC,/home/cpliu/research/papers/d_EDM/PV}

\end{document}